# Wavelet Denoising and Attention-based RNN-ARIMA Model to Predict Forex Price

Zhiwen Zeng
School of Computer Science
The University of Sydney
Sydney, Australia
1910373803@qq.com

Matloob Khushi
School of Computer Science
The University of Sydney
Sydney, Australia
mkhushi@uni.sydney.edu.au

*Abstract*— **Every change of trend in the forex market presents a great opportunity as well as a risk for investors. Accurate forecasting of forex prices is a crucial element in any effective hedging or speculation strategy. However, the complex nature of the forex market makes the predicting problem challenging, which has prompted extensive research from various academic disciplines. In this paper, a novel approach that integrates the wavelet denoising, Attention-based Recurrent Neural Network (ARNN), and Autoregressive Integrated Moving Average (ARIMA) are proposed. Wavelet transform removes the noise from the time series to stabilize the data structure. ARNN model captures the robust and non-linear relationships in the sequence and ARIMA can well fit the linear correlation of the sequential information. By hybridization of the three models, the methodology is capable of modelling dynamic systems such as the forex market. Our experiments on USD/JPY five-minute data outperforms the baseline methods. Root-Mean-Squared-Error (RMSE) of the hybrid approach was found to be 1.65 with a directional accuracy of ~76%.**

## I. INTRODUCTION

Forex stands for foreign exchange is the largest global financial market facilitating daily transactions exceeding $5 trillion [1]. Compared to other financial markets, the decentralized Forex market attracts more industry participants around the world as it allows easier access on 24-hour basis and higher leverage mechanism up to 1:50 [2]. Forex time series forecasting is to mine the potential rule of market movement based on known information to gain profits [3]. Accurate exchange price forecasting is of great value in forex trading and capital investment. Forex market data, and by extension any financial time series, is highly complex and difficult to predict. With investors being influenced by volatile or even chaotic market conditions, financial market trends tend to be non-linear, uncertain and non-stationary.

Great strides in financial data modelling and prediction have occurred over the past 50 years and Autoregressive Integrated Moving Average (ARIMA) is a commonly used one. In the ARIMA theory, one fundamental assumption is that the future value and the historical values of the time series shall satisfy the linear relationship. However, the major financial time series data contain non-linear relationship due to their unstable structure, which limits the scope of the application of ARIMA model.

Recently, miscellaneous machine learning methods especially the Neural Networks (NNs) have achieved promising results in financial forecasting. A survey [4, 5] investigated more than 40 researches on NNs applied in economics and summarized that NNs could discover non-linear relationships in input data making them compatible for modelling non-linear dynamic systems such as forex market. Among all the neural network models, Recurrent Neural Network (RNN) introduces the concept of time series into the design of network architecture, which makes it more adaptable in the analysis of time series data. The attention-based encoder-decoder network is the state-of-art RNN that has shown great potential for sequence prediction in recent works, hence adopted in this paper.

RNN network can detect the non-linear patterns in the sequence, and ARIMA model can well fit the linear relationship in the sequence. By the combination, a novel hybrid methodology can take advantage in both linear and non-linear domains, hence effectively forecasting the complex time series. Due to the highly volatile forex market, the collected time series often contain a great amount of noise, which may mask the true variation of the time series or even change the autocorrelation structure of the sequence. To effectively extract desired information from the original time series, it is necessary to pre-process data to reduce noise via wavelet transform method.

In summary, the proposed system consists of three core methods performing in a synergistic way. Firstly, the Discrete Wavelet Transform (DWT) filters out the disturbance signals in the forex time series. After that, the attention-based RNN and ARIMA models are trained to forecast the non-linear and linear parts of the denoised data respectively. Finally, the two forecasted results are integrated to obtain the final predictions.

The combined model was found to perform better than some traditional NNs in the literature.

## II. RELATED LITERATURE

### A. Wavelet Denoising

A technique to capture cyclicality in original data is wavelet analysis, which transforms a time series or signal into its frequency and time domain [6]. A common utilization of wavelet transform is to reduce data noise. During the past decades, wavelet transform has been proven efficient for data denoising in many application areas such as engineering [7, 8], image processing [9, 10], telecommunication [11], and econometric forecasting [12]. When it comes to the financial data, wavelet transform considers a time series ($f_t$) as a deterministic function ($d_t$) plus random noise ($n_t$) where the $d_t$ contain useful signals and $n_t$ are the interference signals that should be eliminated [13]. Discrete wavelet transform (DWT), a type of wavelet transform with mathematic origin, is appropriate for noise filtering on financial time series [14].

### B. Time Series Analysis

Stochastic statistical models such as Box-Jenkins's Autoregressive Integrated Moving Average (ARIMA) have been proven the potential for short-term trend prediction in time series analysis [15], but their success depend critically on the input data to train the model. That is, when fitting ARIMA models on either non-linear or non-stationary time series, the results of forecasting are expected to be inaccurate, because the predictions always converge to the mean of the series after a few times of forecasting [16]. In real situation, the floating forex data can be highly non-linear because of the market volatility which leads to an undesired ARIMA forecasting results. This motivates an improvement on ARIMA which can stationary the series by releasing non-linear relationship in the original data.

### C. Artificial Neural Networks (ANNs)

Existing works indicated that NNs are able to provide effective means of modelling markets through its flexibility to capture robust and non-linear correlation [17]. The family of Recurrent Neural Networks (RNNs) have recursive feedback linkages between hidden neuron cells forming a directed cycle, which are capable to retain and leverage information from past data to assist the prediction of future values. Such recurrent architectures by nature are tailored for modelling sequential data with delayed temporal correlations. [18] Many recent works with RNNs have shown good promise in econometric price prediction using either technical indicators [19, 20] or social sentiments [21, 22].

Attention-based encoder-decoder network [23] is the state-of-art RNN method that employs an attention mechanism to select important parts of states across all the time steps. Attention mechanisms in NNs are loosely based on the visual attention mechanism found in humans, which essentially come down to being able to focus on certain time steps of an input sequence with "high weight" while perceiving the rest time steps with "low weight", and then adjusting the focal point over time. Over the past four years, the attentioned-RNNs became prevalent for sequence prediction [24, 25], and this paper intends to explore its efficiency in financial analysis.

### D. Hybrid Approach

It has been argued that the hybridization of linear and non-linear models performs better than individuals for time series forecasting. Various types of combing methodologies have been proposed in the literature. [26] introduced a ARIMA-ANN model for time series forecasting and explained the advantage of combination via linear and non-linear domains. They claimed that the ARIMA model fitting contains only the linear component and the residuals contain only the nonlinear behavioral patterns that can be predicted accurately by the ANN model. Rout et. al. [27] implemented adaptive ARIMA models to predict the currency exchange rate finding that the combined models achieved better results. More recently, RNNs which can capture the sequence information were preferred than the simple neural networks to be used in the hybrid models for price predictions [28, 29]. It was also emphasized [30] that the sequential order of combining RNN and ARIMA models impacted on the final predictions and running RNN before ARIMA model provided better accuracy. The same run-time sequence was adopted in this paper.

## III. METHODOLOGIES

In any time-frame of Forex trading there are four prices: open, highest, lowest and close price. We used closed prices of 5 minutes and denoised them using the following wavelet function [31]:

$$X_w(a,b) = \frac{1}{|a|^{\frac{1}{2}}} \int_{-\infty}^{\infty} x(t) \psi_t \left(\frac{t-b}{a}\right) dt \quad (1)$$

Where $\psi_t$ is the continuous mother wavelet which gets scaled by a factor of 'a' and translated by a factor of 'b'. When it comes to the DWT filed, discrete values are used for the scale and translation factors. As the resolution level increases, the scale factor increases in powers of two i.e. a = 1, 2, 4… and the translation factor increases as integers i.e. b = 1, 2, 3…

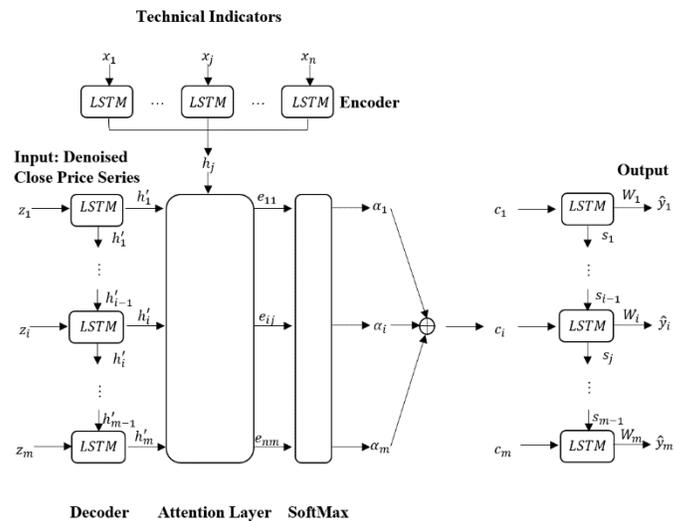

Fig. 1. Graphical illustration of the attention based recurrent neural network.

## A. Autoregressive Integrated Moving Average (ARIMA)

The ARIMA pioneered by Box and Jenkins is a flexible and powerful statistical method for time series forecasting [32]. The ARIMA model considers a time series as a random sequence and approximates the future values as a linear function of the past observations and white noise terms. Basically, the ARIMA consists of three components: 1. Non-seasonal differences for stationarity (I), 2. Auto-regressive model (AR), 3. Moving average model (MA) [33].

To understand the stationary difference order (I), the backward shift operator "B" is introduced, which causes the observation that it multiplies to be shifted backwards in time by 1 period. That is, for any time series R and any period t:

$$BR_t = R_{t-1} \qquad (2)$$

For any integer n, multiplying by B-to-the-nth-power has the effect of shifting an observation backwards by n periods.

$$B^n R_t = B^{n-1}(BR_t) = B^{n-1} R_{t-1} = \cdots = R_{t-n} \qquad (3)$$

Suppose $r_t^d$ denotes for the $d^{th}$ difference lag at time t which has a simple representation in terms of B. Let's start the first-difference operation:

$$r_t^1 = R_t - R_{t-1} = R_t - BR_t = (1-B)R_t \qquad (4)$$

The above equation indicates that the differenced series r is obtained from the original series R by multiplying by a factor of 1-B. Therefore, in general the $d^{th}$ difference $r_t^d$ is given as:

$$r_t^d = (1-B)^d R_t \qquad (5)$$

The linear combination of AR process of order p (AR(P)) and MA model of order q (MA(q)) can be expressed as follows.

$$r_t = c + \varepsilon_t + \sum_{n=1}^{p} \phi_n r_{t-n} = c + \varepsilon_t + \sum_{n=1}^{p} \phi_n B^n r_t \qquad (6)$$

$$r_t = \mu + \varepsilon_t + \sum_{n=1}^{q} \theta_n \varepsilon_{t-n} = \mu + \varepsilon_t + \sum_{n=1}^{1} \theta_n B^n \varepsilon_t \qquad (7)$$

Where the constant p, q are model orders, $\phi_n, \theta_n$ are model parameters, c is a constant, μ is the mean of the series, and $\varepsilon_t \sim WN(0, \sigma^2)$ is the random noise. Considering both AR(P) and MA(q) properties, ARMA (p, q) can be written as:

$$\left(1 - \sum_{n=1}^{p} \phi_n B^n\right) r_t = \left(1 + \sum_{n=1}^{q} \theta_n B^n\right) \varepsilon_t \qquad (8)$$

Combing the above equation with equation (5), the general form of the ARIMA (p, d, q) model can be rewritten as:

$$\phi_p(B)(1-B)^d R_t = \theta_p(B)\varepsilon_t \qquad (9)$$

Where $\phi_p(B) = 1 - \sum_{n=1}^{p} \phi_n B^n$ represents the AR component, $\theta_p(B) = 1 + \sum_{n=1}^{q} \theta_n B^n$ represents the MA component, and d is the number of difference order.

## B. Attention-based Recurrent Neural Network (ARNN)

Attention-based encoder-decoder networks were initially brought out in the field of computer vision and became prevalent in Natural Language Processing (NLP). In this paper, the proposed ARNN follows the structure of a typical encoder-decoder network but with some modifications to perform time series prediction. A graphical illustration of the proposed model is shown in Fig. 1.

Suppose T is the length of window size, for any time t, the n technical indicator series denoised i.e. $X_t = (x_t^1, x_t^2, \ldots, x_t^n)^T = (x_1, x_2, \ldots x_T) \in R^{n \times T}$ are the inputs for encoder, and m close price series i.e. $Z_t = (z_t^1, z_t^2, \ldots, z_t^m)^T = (z_1, z_2, \ldots, z_T) \in R^{m \times T}$ are the exogenous inputs for decoder. Typically, given the future values of the target series (next hour's close price) i.e. $y_t$, the ARNN model aims to learn a non-linear mapping between inputs (X and Z) and target series Y:

$$\hat{y}_t^{ARNN} = f(X_t, Z_t) \qquad (10)$$

Where f is a non-linear mapping function that is a long-short term memory (LSTM). Each LSTM unit has a memory cell with the state $s_t$ at time t, which will be controlled by three sigmoid gates: forget gate $f_t$, input gate $i_t$ and output gate $o_t$. The LSTM unit is updated as follows: [34]

$$f_t = \sigma(W_f[h_{t-1}; x_t] + b_f) \qquad (11)$$

$$i_t = \sigma(W_i[h_{t-1}; x_t] + b_i) \qquad (12)$$

$$o_t = \sigma(W_o[h_{t-1}; x_t] + b_o) \qquad (13)$$

$$s_t = f_t \odot s_{t-1} + i_t + \tanh \odot (W_s[h_{t-1}; x_t] + b_s) \qquad (14)$$

$$h_t = o_t \odot \tanh(s_t) \qquad (15)$$

Where $[h_{t-1}; x_t] \in R^{m+n}$ is a concatenation of the previous hidden state $h_{t-1}$ and the current input $x_t$. $W_f, W_i, W_o, W_s \in R^{m \times (m+n)}$, and $b_f, b_i, b_o, b_s \in R^m$ are parameters to learn. σ and ⊙ are a logistic sigmoid function and an elementwise multiplication, respectively.

*Encoder* is essentially an LSTM that encodes the input sequences (technical indicators) into a feature representation. For time series prediction, given the input sequence $(x_1, x_2, \ldots x_T)$ with $x_j \in R^n$, the encoder can be applied to learn a mapping from $x_t$ to $h_t$ at time step t with

$$h_j = f_1(h_{t-1}, x_j) \qquad (16)$$

Where $h_j \in R^{n_1}$ is the $j^{th}$ hidden state of the encoder, $n_1$ is the size of the hidden state and $f_1$ is a non-linear activation function in a recurrent unit. In this paper, we use stacked two-layer simple RNN as $f_1$ to capture the associations of technical indicators. The mathematic notation for the hidden state update can be formulated as:

$$h_j = \tanh(W_{hh} h_{j-1} + W_{xh} x_j) \qquad (17)$$

Where $W_{hh}$ is the weight matrix based on the previous hidden state and $W_{xh}$ is the weight matrix based on the current input.

*Decoder* use another two-layer LSTM is used to decode the information from denoised close price series i.e. $(z_1, z_2, ..., z_T)$ with $z_i \in R^m$ as:

$$h'_i = f_2(h'_{i-1}, z_i) \quad (18)$$

Where $h'_i \in R^{m_1}$ is the $i^{th}$ hidden state of the decoder, $m_1$ is the size of the hidden state and $f_2$ is a non-linear activation function with the same structure as the $f_1$ in the encoder.

*Attention* mechanism express the $j^{th}$ input of the encoder by a context vector $(c_i)$ as the weighted sum of hidden states that corresponds to the $i^{th}$ output of the decoder.

$$c_i = \sum_{j=1}^{T} \alpha_{ij} h'_i \quad (19)$$

Where $h_j$ is the $j^{th}$ hidden state in the encoder, and $\alpha_{ij}$ is the attention coefficient of sequence obtained from the softmax function:

$$\alpha_{ij} = \frac{\exp(e_{ij})}{\sum_{k=1}^{T} \exp(e_{ik})} \quad (20)$$

Where $e_{ij} = g(s_{i-1}, h_j)$ is called the alignment model, which evaluates the similarity between the $j^{th}$ input of encoder and the $i^{th}$ output of decoder. The dot product is used for the similarity function g in this paper. Given the weighted sum context vector $(c_i)$, the output series of decoder can be computed as:

$$s_i = f_3(h'_i, c_i) \quad (21)$$

Where $h'_i \in R^{m_1}$ is the $i^{th}$ hidden state of the decoder, $s_i$ is the ith output of the decoder and function $f_3$ is chosen as elementwise multiplication in this paper. To predict target $\hat{y}_t$, we use a third LSTM-based RNN on the decoder's output (s):

$$\hat{y}_t^{ARNN} = W^T H[s] + b \quad (22)$$

Where $H[s]$ is one RNN unit, $W^T$ and b are parameters of dense layers that map the RNN neurons to the size of the target output. [35]

### C. Hybrid ARNN-ARIMA Model

In terms of modelling sequence, there are two possible ways to combine the ARNN and ARIMA models. The first method is to use an ARIMA for forest the closing price and an ARNN to predict the residual. The other method is to use ARNN to predict the next hour's closing price and an ARIMA to forecast the residual. This paper adopted the second method as it has been proven suitable for forex data [27]. Fig. 2 shows the high-level flowchart of the hybrid approach.

The ARNN is firstly used to predict the close price $\hat{y}_t^{ARNN}$ in the next hour, then the residual $R_t$ can be calculated as the difference of prediction $\hat{y}_t^{ARNN}$ and ground truth $y_t$.

$$R_t = y_t - \hat{y}_t^{ARNN} \quad (23)$$

This residual series is modelled using an ARIMA model, and the final price $(\hat{y}_t)$ is computed by combining the prediction from ARNN $(\hat{y}_t^{ARNN})$ and residual from ARIMA $(\hat{R}_t)$.

$$\hat{y}_t = \hat{y}_t^{ARNN} + \hat{R}_t \quad (24)$$

IV. EXPERIEMNTS

### A. Data Collection and Pre-processing

The data in the experiment covers 75,000 records of the USDJPY currency pair from 2019-01-01 to 2019-12-31 collected from the MetaTrader5 platform. Figure shows the five-minute (M5) trend of the close price of the obtained data. The last 25% samples are used for testing and the rest are the training set. In the ARNN model, the training set is further split into the training (80%) and validation set (20%) to evaluate the performance meanwhile avoid overfitting.

The raw close price series with technical indicators (listed below) are used as the input for the proposed model.

- Momentum indicators: average directional movement index, absolute price oscillator, arron oscillator, balance of power, commodity channel index, chande momentum oscillator, percentage price oscillator, moving average convergence divergence, williams, momentum, relative strength index, stochastic oscillator, triple exponential average.

- Volatility indicators: average true range, normalized average true range, true range.

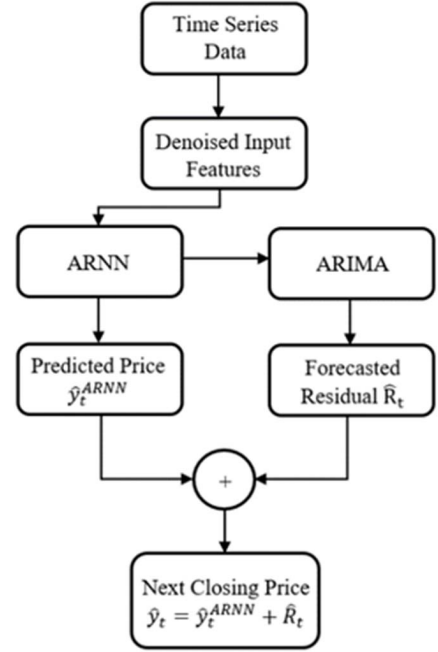

Fig. 2. High level block diagram of the proposed model.

When using the machine learning methods, the original data are usually normalized before modelling to remove the scale effect. In this experiment, Min-Max-scale is conducted on the input data.

$$x_{norm} = \frac{x_t - x_{min}}{x_{max} - x_{min}} \quad (25)$$

Where $x_{norm}$ is the data after normalization, and $x_{min}, x_{max}$ are the minimum and maximum data of the input (X). After modelling the target output are anti-normalized.

$$\hat{y}_t = y_{norm}(y_{max} - y_{min}) + y_{min} \quad (26)$$

Where $\hat{y}_t$ is the predictive price after anti-normalization, $y_{norm}$ is the predictions directly derived from the porposed model, and $y_{max}, y_{min}$ are the minimum and maximum values of the target data (Y).

*B. Performance Evaluation Criteria*

Three evaluation metrics are used to assess the predictive accuracy: (1) the root-mean-squared-error (RMSE), (2) the mean-absolute-percentage-error (MAPE) (3) the directional accuracy. The RMSE is defined as:

$$\text{RMSE} = \sqrt{\frac{\sum_{t=1}^{N}(\hat{y}_t - y_t)^2}{N}} \quad (27)$$

Where $\hat{y}_t$ and $y_t$ are the prediction and ground truth at time t, and N is the number of test samples.

Compared to the RMSE, the MAPE eliminates the influence of the magnitude by using the percentage error, which can be calculated as:

$$\text{MAPE} = \frac{1}{N}\sum_{t=1}^{N}\left|\frac{(\hat{y}_t - y_t)}{y_t}\right| \times 100\% \quad (28)$$

Apparently, the RMSE and MAPE are positive numbers, and the smaller (or closer to 0) the values the higher the accuracy of the model.

In real scenario, the direction of the trend is of significance because traders use the foresting price to place trading orders accordingly. Therefore, we also compute the directional accuracy as following.

$$\text{DA} = \frac{1}{N}\sum_{t=1}^{N} D_t \text{ where } D_t = \begin{cases} 1, (y_{t+1} - y_t)(\hat{y}_{t+1} - y_t) \geq 0 \\ 0, \text{ otherwise} \end{cases} \quad (29)$$

DA is in range [0,1] and the closer DA is to 1, the better the model performs.

*C. Parameter Settings*

In this paper, the wavelet transform function $\psi_t$ was chosen as 'sym15' and the decomposition were conducted for resolution levels up to 4.

For the ARNN model, the number of time steps in the window T is the hyper-parameter to be determined via a grid search over $T \in \{3, 5, 10, 15, 20\}$. When $T = 10$ the model achieved the best accuracy on the validation set. There are three RNNs used in the ARNN models, namely encoder network, decoder network and attention network mapping the context vector to the target. The details of the ARNN parameter setting are summarized in the Table I.

The residual series $R_t$ is obtained as the difference of the ARNN predicted values and actul values, which is expected to account for only the linear and stationary part of the data. For a linear input, the ARIMA model is best fitted to order (p,0,0) where p = 3 was determined by the trial-and-error tests.

The model was structured and trained through the Google's Colaboratory platform with GPU support in the Python3 programming language. The GPU model provided here is the Tesla K80 with pre-installed commonly used frameworks such as TensorFlow. And the hardware space is 2vCPU @ 2.2GHz, 13GB RAM, 33GB Free Space, GPU instance 350 GB.

Table I. Parameters for ARNN model

|  | Encoder | Decoder | Attention |
|---|---|---|---|
| Number of RNN layers | 2 | 2 | 1 |
| Neuron number for RNN layers | (64,32) | (64,32) | (32) |
| Number of Dense layers | 1 | 1 | 2 |
| Neuron number for Dense layers | (64,32) | (64,32) | (16,1) |
| Dimension of input | (20,16) | (20,3) | (10,10) |
| Dimension of output | (10,10) | (10,10) | (None,1) |
| Activation function | ReLu | | |
| Batch size | 64 | | |
| Learning size | 0.001 | | |
| Number of epochs | 100 | | |

Table II. Experimental results

| Architecture | Denoised | RMSE($\times 10^{-3}$) | MAPE (%) | DA (%) | T (minutes) |
|---|---|---|---|---|---|
| RNN | No | 159.58 | 31.1 | 48.2 | 46 |
| RNN | Yes | 2.17 | 34.6 | 62.2 | 48 |
| GRU | No | 98.76 | 16.9 | 48.0 | 95 |
| GRU | Yes | 2.39 | 38.5 | 60.7 | 96 |
| LSTM | No | 88.78 | 16.1 | 48.1 | 102 |
| LSTM | Yes | 2.27 | 34.4 | 63.9 | 116 |
| ARNN | No | 194.63 | 33.7 | 48.8 | 118 |
| ARNN | Yes | 1.71 | 24.2 | 73.5 | 134 |
| ARNN+ARIMA* | Yes | 1.65 | 23.2 | 75.7 | 249 |

DA = Directional Accuracy, GRU = Gated Recurrent Unit, T = Training Time

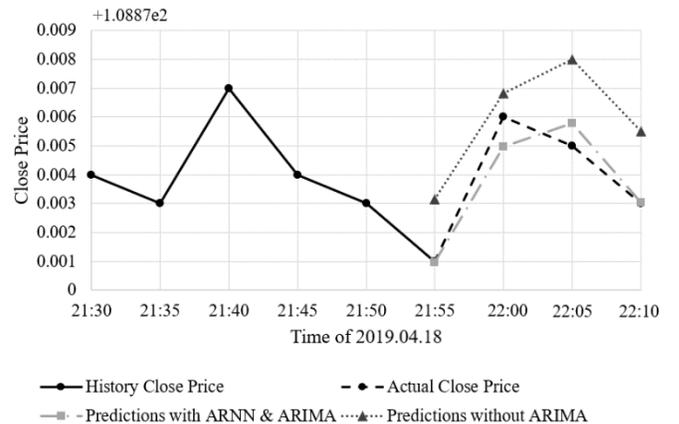

Fig. 3. The effect of the ARIMA model

Table III. Comparison of the proposed model and the literature

|  | SVR | LSTM | ARNN | ARNN+ARIMA* |
|---|---|---|---|---|
| RMSE ($\times 10^{-3}$) | 30.2 | 46.4 | 1.6 | 1.3 |

SVR = Support Vector Regression; * our proposed model

## V. RESULTS AND DISCUSSION

### A. Results of Experiments

Table II. demonstrates the accuracy metrics of the proposed model in comparison to the benchmarks using different network architectures and input features. The original input can either be denoised or not. All benchmarks used the same number of recurrent layers as the proposed model (ARNN+ARIMA), but differed in the type of mapping function (f) as introduced in Section III.B.

From Table II, clearly the attentioned long-short term memory network with lower RMSE, MAPE and larger DA values is superior to the other networks, which confirms the efficiency of applying the encoder-decoder network in financial time series prediction. Combing the ARIMA model, the hybrid approach has outperformed the ARNN with less predictive errors and higher directional accuracy. Fig. 3 shows the effect of ARIMA by enlarging the plots of predictions to a few sample points. The ARIMA helps to reduce the gap between the predictions from ARNN and the actual values, hence improve the model performance. It can also be deduced that denoising data is necessary since models of the same architecture achieved better accuracy when the close price series were denoised.

### B. Comparison to the literature

We also compared the model to the methods published in 2018 to further evaluate its performance. The paper [36] used the same 5-minute USD/JPY data but in a different time period (2017/12/5 to 2018/10/19), therefore, a new experiment was conducted with the same data as [35] to control variable. Table III shows that the proposed method achieved lower root-mean-squared-error (RMSE) and performed better. Given the results, the three methods, namely denoising the original data, adopting the encoder-decoder network, and integrating the neural network with ARIMA model have been proven to improve the accuracy for forex price prediction.

## VI. CONCLUSION

This research proposes a hybrid approach consisting of wavelet denoising, attention-based RNN and ARIMA for predicting the volatile foreign exchange rates. The experimental results in Table II confirms that the integrated system performs better than single recurrent neural networks when applying to the recent data in 2019. Meanwhile, Table IV indicates the superiority of the model by comparing to previously published methods. Although the proposed system achieved good accuracy, there are some limitations of the project:

- Discrete wavelet transform (DWT) that filtered out the white noise with a hard threshold cannot guarantee excellent denoised results. The hard threshold might cause either some useful information being removed or some disturbing noises being reserved. Recent studies shown that wavelet transform with soft thresholds obtained great results for time series prediction. This could be the future direction of the forex analysis.

- The inputs for the encoder network were financial basis technical indicators, which may not fully describe the information in the actual time series. In the future works, another neural network can be used to extract the underlying features from the original data to feed into the encoder.

- Experiments in this paper were performed on USDJPY currency pair. The future work shall involve to exam the proposed model on other currencies.